\documentclass[12pt]{article}

\advance \textheight by \topskip
\makeatletter
\@addtoreset{equation}{section}
  
\makeatother

\makeatletter
\def\bsymb#1{
  \begingroup
  \let\@nomath\@gobble \boldmath
  \mathchoice
    {\hbox{$\m@th\displaystyle#1$}}
    {\hbox{$\m@th\textstyle#1$}}
    {\hbox{$\m@th\scriptstyle#1$}}
    {\hbox{$\m@th\scriptscriptstyle#1$}}
  \endgroup}
\makeatother

\def\cc#1{\kern .7em\hfill #1 \hfill\kern .7em}  
\def\ZZ{\hbox{\it Z\hskip -4.pt Z}}
\def\CC{\hbox{\it l\hskip -5.5pt C\/}}
\def\RR{\hbox{\it I\hskip -2.pt R }}

\newcommand{\nc}{\newcommand}
\nc{\beq}{\begin{equation}}
\nc{\eeq}{\end{equation}}
\nc{\beqa}{\begin{eqnarray}}
\nc{\eeqa}{\end{eqnarray}}
\nc{\noi}{\noindent}

\begin{document}

\title{
{\bf  
Monopole Chern-Simons Term: 
Charge-Monopole System as a Particle with Spin}}

\author{{\sf 
Mikhail S. Plyushchay}\thanks{E-mail: mplyushc@lauca.usach.cl}\\
{\small {\it Departamento de F\'{\i}sica, 
Universidad de Santiago de Chile,
Casilla 307, Santiago 2, Chile}}\\
{\small 
{\it and}}\\
{\small {\it Institute for High Energy Physics, Protvino,
Russia}}}
\date{}

\maketitle
\vskip-1.0cm

\begin{abstract}
The topological nature of Chern-Simons term 
describing the interaction of a charge with magnetic monopole
is manifested in two ways: it
changes the plane dynamical geometry
of a free particle for the cone dynamical geometry
without distorting the free (geodesic) character
of the motion, and in the limit of zero charge's mass it
describes a spin system.
This  observation allows us to
interpret the charge-monopole system alternatively
as a free particle of fixed spin with translational
and spin degrees of freedom interacting via
the helicity constraint, 
or as a symmetric spinning top with dynamical moment
of inertia and ``isospin" U(1) gauge symmetry,
or as a system with higher derivatives.
The last interpretation is used to
get the twistor formulation of the system.
We show that the reparametrization and scale invariant
monopole Chern-Simons term supplied 
with the kinetic term of the same invariance 
gives rise to the alternative description
for the spin, which is related to the 
charge-monopole system in a spherical geometry.
The relationship between the charge-monopole system 
and (2+1)-dimensional anyon is discussed in the light
of the obtained results.
\vskip2mm
\noindent 
\end{abstract}

\newpage

\section{Introduction}

The Dirac charge-monopole system \cite{dir}
was one of the first models with
which the importance of topology for
physics was realized.
The model was investigated in various aspects
classically and quantum mechanically \cite{JackR}-\cite{Jackiwcocyc},
and the Dirac's quantization 
of the charge-monopole constant
was understood naturally in terms 
of the underlying topological fibre bundle structure 
\cite{wy1},\cite{fs}-\cite{path}.
In addition, an interesting
analysis of the Dirac quantization condition relies
on 3-cocycles \cite{Jackiwcocyc}.
Another very seminal interplay of physics and topology
is related to the Chern-Simons (CS) ``effects" 
\cite{chs2}-\cite{Lab},  whose one of the  basic physical
aspects established by Deser, Jackiw and Templeton \cite{chs2}
consists in quantization of coupling in field 
theory analogous to Dirac quantization in quantum mechanics.

The spin particle nature of the charge-monopole 
system was observed by Jackiw, Rebbi,
Hasenfrantz and  't Hooft  in the context
of the ``spin from isospin" field theoretical 
mechanism \cite{JackR,HastH},
but the first results on the nature of this 
system anticipating its interpretation as a
particle with spin were obtained  
by Poincar\'e (see ref. \cite{godoli}).
In this paper we exploit the CS nature 
of the charge-monopole coupling term 
to interpret the charge-monopole system
as a particle with spin from the opposite side,
not appealing to its field theoretical origin
but treating it classically and quantum mechanically
as a system with finite number of degrees of freedom. 
More specifically, we observe that the charge-monopole
coupling term having a nature of (0+1)-dimensional
CS term manifests its topological nature
in two ways. First, the  CS term changes the plane dynamical geometry
of a free particle for the cone dynamical geometry
of a charged particle without distorting the free 
(geodesic) character of the motion.
Second, in the limit of zero charge's mass 
the CS term describes a spin system.
This observation allows us to interpret the charge-monopole
system alternatively:

\noindent
({\bf i}) as a free particle of fixed spin with translational and spin degrees
of freedom interacting via the helicity constraint;
in such interpretation, noncommuting gauge-covariant
momenta (velocities) are the analogs of the Foldy-Wouthuysen
coordinates of the Dirac particle;

\noindent
({\bf ii}) as a symmetric spinning top with dynamical 
moment of inertia and ``isospin" U(1) gauge symmetry;

\noindent
({\bf iii}) as a system with higher derivatives; this interpretation
is the charge-monopole analog of another known classical equivalence 
\cite{tor4}
between the relativistic scalar massive particle in a background
of the constant homogeneous electromagnetic field
and higher derivative model of relativistic particle with torsion \cite{Pol}
underlying (2+1)-dimensional anyons \cite{any0}-\cite{tor5}; 

\noindent
({\bf iv}) as an analog of the massless particle with nonzero spin,
that naturally leads to the twistor formulation
for the charge-monopole system.

\noindent
We also show that the reparametrization and scale invariant
CS term supplied with the kinetic term of the same invariance
gives rise to the alternative description for the spin. 
The partially gauge fixed version of such spin model 
corresponds to the charge-monopole system in a spherical
geometry. The obtained results are used to discuss
the relationship between the
charge-monopole system and (2+1)-dimensional anyon.

The paper is organized as follows.
Section 2 is devoted to discussion of the classical
theory of the charge-monopole system in the 
context of its similarity to the 3D free particle.
Here we observe the geodesic
character of the charge's motion and 
demonstrate that being reduced 
to the fixed level of the 
angular momentum integral,
it is described by the Lagrangian
corresponding to the 2D non-relativistic 
particle in the planar gravitational field
of a point-like source \cite{genqm}
carrying simultaneously a nontrivial magnetic flux.
This, in particular, explains  
why the charge-monopole system
and 2D charged particle moving in the field
of the point magnetic vortex have the same  ``hidden"
or ``dynamical" SO(2,1) symmetry 
revealed by Jackiw \cite{so2,jackvort}. 
We obtain the general solution
to the canonical equations of motion, 
compare the charge-monopole
system with the 3D free particle from the point of view
of integrals of motion and their Lie-Poisson
algebras, and, finally, interpret the charge-monopole
system as a reduced E(3) system.
In section 3 we discuss the classical and quantum theory
of the spin represented in the form of 
the (0+1)-dimensional topological theory
given by the CS charge-monopole action corresponding to the
limit of zero charge's mass.
In section 4 we construct the description 
of the charge-monopole system as a 
particle with spin.
We start with the free particle system with spin,
whose value is fixed by the charge-monopole constant.
Then we switch on interaction between translational
and spin degrees of freedom by introducing 
the helicity constraint which freezes
spin degrees of freedom 
and provides finally the physical equivalence
of the extended model to the initial charge-monopole
system.
In section 5 the system is interpreted 
as a spinning symmetric top. In this picture the initial
U(1) electromagnetic gauge symmetry
is changed for the U(1) gauge symmetry
generated by the ``isospin"-fixing constraint,
which makes the rotations about the 
top's symmetry axis to be unobservable.
The spinning top picture is used in section 6
to get the higher derivative form for
the CS term, which, in turn, is
employed in section 7 for constructing
the twistor formulation for the charge-monopole system
proceeding from its analogy to the 
4D massless particle with spin.
In section 8 we discuss the charge-monopole system
in a spherical geometry and find the alternative
formulation for the spin. 
Here we also observe that 
the SO(2,1) symmetry of the charge-monopole system
\cite{so2,jackvort} can be treated formally as a relic of 
the reparametrization invariance surviving 
the Lagrangian gauge fixing procedure
applied to the Euclidean relativistic version
of the model.
Section 9 contains discussion of the relationship between
the charge-monopole system and (2+1)-dimensional 
anyon, and in the last section we present some
concluding remarks.

\section{Charge-monopole dynamics and 3D free particle}
\subsection{Lagrangian formalism} 
A non-relativistic particle of unit mass and electric charge
$e$
in the field of magnetic monopole of charge $g$ is described
by the Lagrangian
\beq
\label{lchmo}
L=\frac{1}{2}{\dot{\bf r}}{}^2+e{\bf A}{\dot{\bf r}},
\eeq
with a U(1)  gauge potential ${\bf A}({\bf r})$ 
defined by the relations
\beq
\partial_iA_j-\partial_jA_i= F_{ij}=\epsilon_{ijk}B_k,\quad
B_i=g\frac{r_i}{r^3},\quad
r=\sqrt{{\bf r}^2}.
\label{lmono}
\eeq
The case of arbitrary mass $m$ 
can be obtained from (\ref{lmono}) 
via the transformation of
time and charge: $t\rightarrow m^{-1/2}t$,
$e\rightarrow m^{1/2}e$.
In definition (\ref{lmono}) it is assumed that
the point $r=0$ is excluded, i.e.
the configuration space of the system,
${\cal M}=\RR^3-\{0\}$, is diffeomorphic to 
$(0,\infty)\times S^2$ and inherits a nontrivial
topology of a two-sphere $S^2$.
It is well known that the electromagnetic potential
(\ref{lmono}) gives a connection of the 
monopole U(1) fibre bundle being a nontrivial Hopf bundle
over $S^2$, and the problems with Dirac strings of singularity
can be escaped by covering the configuration space 
${\cal M}$ with two charts \cite{topology,Nakahara}.
These topological complications, however,
do not play any role for us under treating the
classical dynamics of the system,
but will reveal themselves at the quantum level.

Equations of motion following from Lagrangian (\ref{lchmo})
result in the Lorentz force law,
\beq
\label{chmr}
\ddot{\bf r}=-\frac{\nu}{r^3}\cdot{\bf r}\times {\dot{\bf r}},\quad
\nu=eg,
\eeq
which implies that instead of the orbital angular momentum vector
$
{\bf L}={\bf r}\times {\dot{\bf r}}, 
$
the vector 
\beq
\label{jint}
{\bf J}={\bf L}-\nu{\bf n},
\quad {\bf n}={\bf r}\cdot {r^{-1}},
\eeq
is the conserved 
angular momentum of the system.
The nontrivial term $-\nu{\bf n}$
can be understood as the electromagnetic
angular momentum produced by both
the electric charge and magnetic monopole (see ref. \cite{godoli}).
Note that $J=\sqrt{{\bf J}^2}$ is restricted from below
by the modulus of the charge-monopole coupling constant 
$\nu$, $J\geq |\nu|$.
Due to the relation  ${\bf Jn}=-\nu$,
the trajectory of the particle 
lies on the cone.
The cone's axis is given by the
vector ${\bf J}$ and its
half-angle is 
\beq
\label{hang}
\cos\gamma=-\nu J^{-1}.
\eeq
Since the force ${\bf f}=-\nu r^{-3}{\bf r}\times
\dot{\bf r}$ is orthogonal to
${\bf r}$ and to the velocity $\dot{\bf r}$,
it is perpendicular to the cone. Therefore,
the particle performs a  {\it free motion on the cone}.
This can also be observed directly as follows.
Taking into account the conservation of the 
angular momentum ${\bf J}$ and that 
in spherical coordinates  
the charge-monopole coupling term
takes a form $L_{int}=\nu\cos\vartheta\dot{\varphi}$
(see Eq. (\ref{lspinsing})), we can choose the system of coordinates
with axis $\vartheta=0$ directed along the vector ${\bf J}$.
Then in accordance with Eq. (\ref{hang}), 
$\cos\vartheta=-\nu J^{-1}$, $\dot{\vartheta}=0$,
and Lagrangian (\ref{lchmo}) is reduced to
$L=\frac{1}{2}(\dot{r}{}^2+(1-\nu^2 J^{-2})r^2\dot{\varphi}{}^2)
-\nu^{2}J^{-1}\dot{\varphi}$.
After transformation 
\beq
\label{rescal}
t\rightarrow  t'=\alpha t,\quad \alpha=\sqrt{1-\nu^2 J^{-2}},
\eeq
we obtain finally the following form for the 
charge-monopole Lagrangian:
\beq
\label{conus}
L=\frac{1}{2}(\alpha^{-2}\dot{r}{}^2+r^2\dot{\varphi})
-\alpha \nu^2 J^{-1}\dot{\varphi}.
\eeq
The last term is the total time derivative of 
the topologically nontrivial angular variable.
It is the reduced form of the 
charge-monopole interaction term having a nature
of $(0+1)$-dimensional CS term \cite{djt}-\cite{asorey}.
As we shall see,
quantum mechanically this will give rise to
the Dirac quantization condition for $\nu$,
but classically the total derivative can be omitted
without changing the equations of motion.
The topologically nontrivial term corresponds exactly 
to the 2D term describing the interaction 
of the charge $e$ with a (singular) point vortex
carrying the magnetic flux \cite{jackvort}
$
\Phi=-2\pi \alpha \nu^2 J^{-1}e^{-1}.
$
Without the last term, Eq. 
(\ref{conus}) is a Lagrangian of a 
{\it free} particle of unit mass on the cone
given by the relations
$
x=r\cos\varphi,
$
$
y=r\sin\varphi,
$
$
z=r\sqrt{\alpha^{-2}-1},
$
$r>0,
$ 
$
0\leq \varphi\leq 2\pi,
$
with $0<\alpha<1$,
that confirms our statement on a free (geodesic)
motion of the charge over the cone  (\ref{hang}).
Together with the pointed out nature of the total derivative 
term, 
this, as we shall see, explains why the charge-monopole system
and 2D charged particle moving in the field
of the point magnetic vortex have the same  
dynamical SO(2,1) symmetry \cite{so2,jackvort}. 

Since the conical metric
$ds^2=\alpha^{-2}(dr)^2+r^2(d\varphi)^2$
corresponds to the metric produced by the point mass \cite{genqm,Mor},
we can say that the classical motion of the charge in 
the field of magnetic monopole (reduced 
to the fixed value of the integral
${\bf J}$) is equivalent to the classical motion 
of a particle in a planar gravitational 
field of a point massive source
carrying simultaneously magnetic flux 
$
\Phi=-2\pi \alpha \nu^2 J^{-1}e^{-1}.
$

{}From the form of transformation (\ref{rescal})
and Lagrangian (\ref{conus}) it is clear that
the case $J=|\nu|$ is singular
and should be treated as a limit case,
i.e. we have to assume that $J>|\nu|$.
We shall discuss this peculiarity of the charge-monopole
system in different aspects in what follows.

In the case of a 3D free particle ($\nu=0$, ${\cal M}=\RR^3$),
the motion is characterized 
by the coordinate ${\bf r}$ 
and by the conserved linear momentum ${\bf p}$.
Alternatively, with the appropriate
choice of the origin of the system of coordinates,
the particle's motion can be characterized by
the unit vector ${\bf n}$ and by the 
conserved orbital angular momentum ${\bf L}$ 
supplemented with the canonically conjugate scalars $r$ and 
$p_r={\bf pr}\cdot r^{-1}$.
Since ${\bf Ln}=0$, for a given ${\bf L}$
the particle's trajectory is in  the plane
orthogonal to the orbital angular momentum.
So, we conclude that classically the topological nature 
of $(0+1)$-dimensional charge-monopole CS term
is manifested in changing the 
global structure of the dynamics without distorting 
its local free (geodesic) character:
the ``plane dynamical geometry" of the free particle ($e=0$)
is changed for the free ``cone dynamical geometry" of the charged
particle. 
We shall discuss the relation between the two systems 
in the context of (dynamical) integrals of motion in 
subsection 2.3.

\subsection{Canonical  formalism: solutions to the equations
of motion}

To solve the equations of motion in general form and analyze
in more detail the system's dynamics,
we turn to the canonical formalism.
The Hamiltonian corresponding to Lagrangian
(\ref{lmono}) is 
$
H=\frac{1}{2}{\bf {P}}^2,
$
where 
$
{\bf P}={\bf p}-e{\bf A}
$
is a classical analog of the gauge-covariant 
derivative defined via the
momentum ${\bf p}$ canonically conjugate to ${\bf r}$.
This gives rise
to the Poisson brackets
\beq
\label{brmono}
\{P_i,P_j\}=\frac{\nu}{r^3}\epsilon_{ijk}{r_k},\quad
\{r_i,P_j\}=\delta_{ij},\quad
\{r_i,r_j\}=0,
\eeq 
and to the equations of motion
\beq
\label{eom}
\dot{\bf r}={\bf P},\quad
\dot{\bf P}=-\frac{\nu}{r^3}{\bf L},
\eeq
with ${\bf L}={\bf r}\times {\bf P}$.
{}From Eq. (\ref{eom}) we find
that the vectors ${\bf n}$
and ${\bf L}$ evolve according to the equations
$
\dot{\bf L}=\nu r^{-2}\cdot {\bf L}\times {\bf n},
$
$
\dot{\bf n}=r^{-2}\cdot {\bf L}\times {\bf n},
$
and, as a consequence, precess about the integral ${\bf J}={\bf L}-\nu {\bf n}$
with the same frequency:
$
\dot{\bf L}=r^{-2}\cdot{\bf J}\times {\bf L},
$
$
\dot{\bf n}=r^{-2}\cdot{\bf J}\times {\bf n}.
$
The vectors
${\bf J}$ and ${\bf n}$ obeying the relations
${\bf n}^2=1$ and ${\bf Jn}=-\nu$, 
together with the two scalars 
${\bf r}^2$ and ${\bf Pr}$
form the complete set of observables
in terms of which ${\bf r}$ and ${\bf P}$
can be completely ``restored".
The equations of motion for the scalar observables
have a simple form
$
({\bf r}{}^2)\dot{}=2{\bf Pr},
$
$
({\bf Pr})\dot{}={\bf P}^2=2H,
$
being exactly the same as in the case of a free particle.
Their integration gives
\beq
\label{pr}
({\bf Pr})(t)=({\bf Pr})(t_0)+{\bf P}^2\cdot(t-t_0),\quad
{\bf r}^2(t)={\bf P}^2\cdot(t-t_0)^2+2({\bf Pr})(t_0)\cdot
(t-t_0)+{\bf r}^2(t_0).
\eeq
The minimal charge-monopole distance
corresponds to the moment of time for which
${\bf Pr}=0$, and is given by the relation
${\bf r}^2_{\min}={\bf L}^2\cdot {\bf P}^{-2}$.
To complete the integration of equations of motion, we  
solve the equation 
$
\dot{\bf n}=r^{-2}\cdot{\bf J}\times {\bf n}
$
and get
\beqa
&{\bf n}(t)=-\nu J^{-1}{\bf j}
+{\bf n}_\perp(t),\quad
{\bf n}_\perp(t)=\left({\bf n}(t_0)+
 {\bf j}\nu J^{-1}\right)\cos \tau(t)
 + {\bf j}\times{\bf n}(t_0)\sin \tau(t),&
\label{nt}\\ 
&\tau=JL^{-1}\cdot \tan^{-1} \left({\bf Pr}\cdot L^{-1}\right),\quad
L=\sqrt{{\bf J}^2-\nu^2},&\label{tau}
\eeqa
where ${\bf j}={\bf J}\cdot J^{-1}$, 
and evolution of ${\bf Pr}$ is given by Eq. (\ref{pr}). 
Zero value of the angular function $\tau(t)$ corresponds 
to the point of perihelion ($r=r_{min}$) with respect to which
the trajectory is symmetric.
The vector ${\bf n}_\perp$ is the projection of 
${\bf n}$ to the plane orthogonal to the 
angular momentum ${\bf J}$,
and Eqs. (\ref{nt}), (\ref{tau}) give the 
classical scattering angle of the particle's motion
projected into the plane perpendicular to ${\bf J}$
as a function of it:
\beq
\label{scat}
\varphi_{scat}^\perp=\tau(+\infty)-\tau(-\infty)=\pi JL^{-1}.
\eeq
Eq. (\ref{nt}) together with Eq. (\ref{pr}) and relation
${\bf P}=\dot{\bf r}$ give a complete solution to the equations of motion
(\ref{eom}).

In correspondence with Eqs. (\ref{hang}), (\ref{scat}),
in the limit $J\rightarrow \infty$
the cone over which the particle moves
is close to the plane:
$\gamma\rightarrow \pi/2$
and $\varphi_{scat}^\perp \rightarrow \pi$.
In another limit, $J\rightarrow \vert \nu\vert$,
the cone is degenerated into a half-line, 
$\gamma\rightarrow 0 (\pi)$ for $\nu<0(>0)$,
whereas the  number of full rotations
${\cal N}=[\varphi_{scat}^\perp/2\pi]$ is infinite, 
where $[.]$ is the integer part.
Therefore, the case $J=|\nu|$ (${\bf L}=0$) with ${\bf P}\neq0$,
which corresponds to the motion of the charge
with constant velocity along a straight half-line defined by ${\bf n}(t_0)$
in the direction to or from  the monopole 
is not a proper limit:
such a trajectory corresponds
to the half of the trajectory with $J\rightarrow\infty$
($L\rightarrow\infty$) but not to 
the limit $J\rightarrow|\nu|$.
This supports  our conclusion of the previous subsection
on a necessity to treat the values of the angular momentum
to be confined to the domain $|\nu|<J<\infty$.

\subsection{Integrals of motion and their algebra}

Let us compare the charge-monopole system with a 3D free particle
from the point of view of integrals of motion 
and corresponding Lie-Poisson algebras formed by them.

For the $3D$ free particle ($\nu=0$),
the first integrals of motion are linear, ${\bf p}$, and angular,  
${\bf L}={\bf r}\times{\bf p}$, momenta forming the set of 
5 (${\bf pL}=0$) algebraically independent conserved quantities
not depending explicitly on time.
With respect to the Poisson brackets, they  
form the algebra of Euclidean group $E(3)$.
Algebraically, the Hamiltonian is a square of the vector ${\bf p}$,
$H=\frac{1}{2}{\bf p}^2$,
but it generates  the symmetry of time translations  being independent
from the symmetries of space translations and rotations
generated by ${\bf p}$ and ${\bf L}$. 
This is not the only independent symmetry 
which can be generated via constructing 
algebraically dependent quantities from ${\bf L}$ and ${\bf p}$. 
Indeed, the vector ${\bf Q}={\bf p}\times {\bf L}$ is the analog of the 
Laplace-Runge-Lenz vector of the Coulomb-Kepler system, which can be
treated as a generator of the corresponding canonical symmetry transformations.
The integrals  ${\bf L}$ and ${\bf Q}$, ${\bf LQ}=0$, form together 
with the Hamiltonian the nonlinear algebra:
\[
\{L_i,V_j\}=\epsilon_{ijk}V_k,\quad
\{Q_i,Q_j\}=-2H\epsilon_{ijk}L_k,\quad
\{H,V_i\}=0,\quad
V_i=L_i,Q_i.
\]
The renormalized (at ${\bf p}\neq 0$)
vector ${\bf R}\equiv {\bf Q}/\sqrt{{\bf p}^2}$ 
and the angular momentum vector ${\bf L}$
form the Lorentz algebra $so(3,1)$.
One can construct another vector, 
${\bf K}\equiv {\bf L}\times{\bf R}\cdot L^{-1}$,
$L=\sqrt{{\bf L}^2}$,
which together with ${\bf R}$
and ${\bf L}$ provides us with the complete set 
of the three orthogonal vectors,
${\bf RL}={\bf KL}={\bf RK}=0$, 
of the same norm,
${\bf R}^2={\bf K}^2={\bf L}^2$.
They  form the following nonlinear algebra:
\beq
\{L_i,V_j\}=\epsilon_{ijk}V_k,\quad
\{R_i,R_j\}=\{K_i,K_j\}=-\epsilon_{ijk}L_k,\quad
\{R_i,K_j\}=\delta_{ij}L,
\label{rlk2}
\eeq
where $V_i=L_i,R_i,K_i$.
One notes that
the scalar $L$ rotates the set of
vectors ${\bf R}$ and ${\bf K}$:
\beq
\label{lrk}
\{L,R_i\}=-K_i,\quad
\{L,K_i\}=R_i.
\eeq
Therefore, the vector integral ${\bf K}$ 
also forms the $so(3,1)$ algebra with the orbital angular momentum
and is (non-canonically)
conjugate 
to ${\bf R}$ (see
the last Poisson bracket relation in
Eq. (\ref{rlk2})).
The rotated about ${\bf L}$ vectors
${\bf R}'={\bf R}\cdot
\cos \varphi +{\bf K}\cdot\sin \varphi$ and
${\bf K}'={\bf K}\cdot\cos\varphi-
{\bf R}\cdot\sin\varphi$, $\varphi=const$, 
possess exactly the same set of properties
as ${\bf R}$ and ${\bf K}$.

There are also the so called dynamical integrals of motion \cite{so2}
depending explicitly 
on time and they appear as follows.
Solving the equations of motion
$\dot{\bf r}={\bf p}$, $\dot{\bf p}=0$,
one gets 
$
{\bf r}={\bf p}\cdot (t-t_0)+{\bf X},$
${\bf X}\equiv{\bf r}(t_0).$
One can treat ${\bf X}={\bf r}-{\bf p}\cdot(t-t_0)$
as a vector integral of motion
dependent explicitly on time:
$\frac{d}{dt}{\bf X}=\{{\bf X},H\}+\partial {\bf X}/\partial t=0$.
It generates the transformations $x_i\rightarrow x_i -\varepsilon_i(t-t_0)$
corresponding to the Galilei boosts.
The integrals of motion ${\bf p}$ and ${\bf X}$ 
satisfy the Heisenberg algebra $\{X_i,X_j\}=
\{p_i,p_j\}=0$, $\{X_i,p_j\}=1\cdot \delta_{ij}$.
Algebraically, the vector ${\bf X}$ is equivalent to the vector integral
not containing the explicit dependence on time,
$
{\bf X}_\bot=
{\bf r}-{\bf p}{({\bf pr})}\cdot{\bf p}^{-2}={\bf Q}\cdot {\bf p}^{-2},
$
${\bf X}_\bot{\bf p}=0,
$
and to the scalar 
$
D={\bf Xp}={\bf rp}-{\bf p}^2\cdot (t-t_0)
$
being a dynamical integral of motion generating the time dilations \cite{so2,jackvort}.
One notes that the angular momentum ${\bf L}$ 
can also be treated as an integral
algebraically dependent on ${\bf X}$ 
and ${\bf p}$:
${\bf L}={\bf X}\times {\bf p}={\bf X}_\bot\times {\bf p}$.
On the other hand, from the equations of motion 
it follows that 
$
\frac{d}{dt}{\bf r}^2=2{\bf pr},
$
$
\frac{d}{dt}({\bf pr})={\bf p}^2.
$
Integration of the second equation gives rise to 
the dynamical integral $D$ and 
the subsequent integration of the first equation
gives, similar to Eq. (\ref{pr}),
${\bf r}^2(t)={\bf p}^2\cdot (t~-~t_0)^2+2D\cdot (t-t_0)+{\bf r}^2(t_0)$.
The last relation can be rewritten in the form of the
dynamical integral of motion 
$
{\cal R}={\bf r}^2(t_0)={\bf r}^2-{\bf p}^2\cdot (t-t_0)^2-2D\cdot (t-t_0),
$
which generates the time special conformal transformations \cite{so2,jackvort}.
It can be represented equivalently as
\beq
\label{rldp}
{\cal R}=({\bf L}^2+D^2)\cdot {\bf p}^{-2}.
\eeq
The integrals $D$, $2H$ and ${\cal R}$ form the 
same algebra as the scalars ${\bf pr}$, ${\bf p}^2$ and ${\bf r}^2$
(the latter set is reduced to these integrals at the initial moment
$t=t_0$), which is the $so(2,1)\sim sl(2,R)$ algebra
\cite{r-2aff}-\cite{papad}:
\beq
\label{so21}
\{
{\cal J}_0,{\cal J}_1\}={\cal J}_2,\quad
\{
{\cal J}_0,{\cal J}_2\}=-{\cal J}_1,\quad
\{
{\cal J}_1,{\cal J}_2\}=-{\cal J}_0,
\eeq
where ${\cal J}_0=\frac{1}{4}(2H+{\cal R})$,
${\cal J}_1=\frac{1}{4}(2H-{\cal R})$,
${\cal J}_2=\frac{1}{2}D$,
with the Casimir central element 
${\cal C}=-{\cal J}_0^2+{\cal J}_1^2+{\cal J}_2^2=
-\frac{1}{4}{\bf L}^2\leq 0$.
The dynamical integral $D$ together with $H$
commutes in the sense of Poisson brackets 
with the vector integrals ${\bf L}$,
${\bf R}$ and ${\bf K}$,
whereas the integral ${\cal R}$
due to Eqs. (\ref{lrk}), (\ref{rldp})
has nontrivial Poisson bracket 
relations with ${\bf R}$ and ${\bf K}$.

Let us turn to the charge-monopole system,
where instead of the orbital angular momentum, the 
vector ${\bf J}={\bf L}-\nu{\bf n}$ is 
conserved.
Since the equations of motion for 
the scalar variables ${\bf r}^2$ and 
${\bf Pr}$ look exactly as the corresponding equations
for the free particle (with the change of ${\bf p}$
for ${\bf P}$), the charge-monopole system
has the set of the scalar dynamical integrals 
of motion of the same form \cite{so2},
\beq
D={\bf Pr}-{\bf P}^2\cdot(t-t_0),
\label{D}\quad
{\cal R}={\bf r}^2-{\bf P}^2\cdot (t-t_0)^2-2D\cdot(t-t_0).
\label{R}
\eeq
Though $P_i$ are characterized by the 
nontrivial Poisson brackets (\ref{brmono}),
the scalar dynamical integrals (\ref{R}) and 
the Hamiltonian 
generate the same $so(2,1)$ algebra (\ref{so21})
as in a free case with the Casimir central element
\beq
\label{casim}
{\cal C}=-{\cal J}_0^2+{\cal J}_1^2+{\cal J}_2^2=
-\frac{1}{4}({\bf J}^2-\nu^2)<0.
\eeq
The free particle's vector integrals
of motion ${\bf R}$ and ${\bf K}$
also have analogs in the charge-monopole system.
These are given by the vectors
\beq
\label{rlmono}
{\bf R}\equiv {\bf N}\cdot \frac{J^2}{\sqrt{{J}^2-\nu^2}},
\quad
{\bf K}\equiv{\bf J}\times {\bf N}\cdot\frac{J}{\sqrt{{J}^2-
\nu^2}},
\eeq
where the vector integral
$
{\bf N}=\left({\bf n}+\nu J^{-1}\cdot {\bf j}\right)
\cdot\cos\tau
-{\bf j}\times{\bf n}\cdot\sin\tau
$
can be identified with the vector ${\bf n}_\bot(t_0)$
(see Eq. (\ref{nt})).
Vector ${\bf N}$ satisfies the relations
${\bf NJ}=0$, ${\bf N}^2=1-\nu^2 J^{-2}$,
and 
$
\{N_i,N_j\}=-{\nu^2}J^{-4}\cdot
\epsilon_{ijk}J_k,
$
$
\{N_i,{\bf r}^2\}=\{N_i,{\bf Pr}\}=\{N_i,{\bf P}^2\}=0.
$
{}From these relations we find that
${\bf RJ}={\bf KJ}={\bf RK}=0$,
${\bf R}^2={\bf K}^2={\bf J}^2$,
and that the Poisson bracket algebra of the
complete set of orthogonal vectors
${\bf J}$, ${\bf R}$ and ${\bf K}$
is given by Eq. (\ref{rlk2})
with ${\bf L}$ changed for ${\bf J}$. 
This set of vector integrals is in involution with
the dynamical scalar integral of motion
$D$ and with $H$,
but the vectors ${\bf R}$ and ${\bf K}$
have nontrivial Poisson bracket relations 
with the dynamical integral 
${\cal R}=(D^2+{\bf J}^2-\nu^2)/2H$
(see Eq.  (\ref{lrk})).

Therefore, the dynamical geometry 
similarity of the charge-monopole
system to the 3D free particle discussed in subsection 2.1
also reveals itself in existence of similar
sets of integrals of motion 
(depending and not depending explicitly on time),
which form between themselves the same (nonlinear) Lie-Poisson algebras.

\subsection{Charge-monopole as a reduced E(3) system}
Like a 3D free particle, 
the charge-monopole system may be treated
as a reduced E(3) system.
To get such an interpretation,
let us pass over from the  Hamiltonian variables ${\bf r}$ and ${\bf P}$
to the set of variables ${\bf n}$, ${\bf J}$,
$r$ and $P_r={\bf Pr}\cdot r^{-1}$.
They have the following Poisson brackets:
\beqa
&\{r,P_r\}=1,\quad
\{r,{\bf n}\}=\{r,{\bf J}\}=\{P_r,{\bf n}\}=\{P_r,{\bf J}\}=0,&
\label{0Poi}\\
&\{J_i,J_j\}=\epsilon_{ijk}J_k,\quad
\{J_i,n_j\}=\epsilon_{ijk}n_k,\quad
\{n_i,n_j\}=0.&
\label{njPoi}
\eeqa
Poisson brackets (\ref{njPoi}) correspond to the 
algebra of generators of the Euclidean group E(3) with
$J_i$ being a set of generators of rotations
and $n_i$ identified as generators
of translations. The quantities ${\bf n}^2$ and 
${\bf Jn}$ lying in the center
of $e(3)$ algebra, $\{{\bf n}^2,n_i\}=\{{\bf n}^2,J_i\}=
\{{\bf nJ},n_i\}=\{{\bf nJ},J_i\}=0$,
are fixed in the present case by the relations
\beq
\label{centr}
{\bf n}^2=1,\quad 
{\bf nJ}=-\nu.
\eeq
In terms of the introduced variables,
the Hamiltonian of the system takes the form
\beq
\label{hame3}
H=\frac{1}{2}P_r^2+\frac{({\bf J}\times {\bf n})^2}{2r^2}.
\eeq
Therefore, the charge-monopole system can be treated
as the E(3) system reduced by the conditions 
(\ref{centr}) fixing the Casimir elements 
and supplemented by the independent canonically
conjugate variables $r$ and $P_r$.
It is the second relation from Eq. (\ref{centr})
that encodes the topological difference between
the charge-monopole and the 3D free particle cases:
for $\nu\neq0$, the space given by the spin
vector ${\bf J}$ is homeomorphic to
$\RR^3-\{0\}$, ($J>|\nu|$),
whereas for $\nu=0$ the corresponding space 
$\RR^3$ is topologically trivial.
In the next section we shall discuss the physical consequences
of the nontrivial topological structure 
of the charge-monopole system.

In accordance with Eqs. (\ref{centr}),
one could treat the vector ${\bf L}={\bf J}+\nu{\bf n}$,
${\bf Ln}=0$, as an orbital 
angular momentum.
However,  the Poisson bracket relations $\{L_i,L_j\}=
\epsilon_{ijk}(L_k+\nu n_k)$ 
following from (\ref{njPoi}) 
and restriction ${\bf L}^2>0$ corresponding to
$J>|\nu|$ prevent such interpretation.
Nevertheless, as we shall see, 
it is possible to treat ${\bf L}$ as an orbital angular momentum
in extended physically equivalent 
formulation of the model.

\section{Monopole Chern-Simons term and spin}

This section contains mainly the known
results which are necessary for the self-contained 
presentation of the subsequent analysis.

The integrand in action corresponding to the
the charge-monopole interaction term,
$\theta=e\dot{\bf r}{\bf A}({\bf r})dt$,
can be treated as a differential one-form,
$
\theta=e{\bf A}({\bf r})d{\bf r}.
$
Then the relations (\ref{lmono}) defining the monopole vector potential
are equivalent to the relation
\beq
\label{ch-s}
d\theta=\frac{\nu}{2r^3}\cdot \epsilon_{ijk}r_idr_j\wedge dr_k.
\eeq
The right-hand side of Eq. (\ref{ch-s}) is 
the gauge-invariant
curvature two-form, 
\beq
\label{curvat}
d\theta=e{\cal F},\quad 
{\cal F}=\frac{1}{2}F_{ij}dr_i\wedge dr_j
\eeq
and, consequently,
the gauge-non-invariant one-form $\theta$
has a sense of $(0+1)$-dimensional CS term 
\cite{djt}-\cite{Nakahara}.
The two-form (\ref{ch-s}) can be represented equivalently as
\beq
\label{ds2}
d\theta=\frac{\nu}{2}\epsilon_{ijk}n_idn_j\wedge dn_k,
\quad n_i=r_i\cdot r^{-1}.
\eeq
Via the (local) parametrization by the spherical angles,
${\bf n}={\bf n}(\vartheta,\varphi)$, this
can be treated as the differential area of a two-sphere multiplied 
by the charge-monopole coupling constant:
$d\theta=\nu d(\cos\vartheta)\wedge d\varphi$.
If the vector ${\bf n}(t)$ describes some
closed curve $\Gamma$ on the sphere (i.e. if ${\bf n}(t_1)=
{\bf n}(t_2)$)\footnote{
According to Eqs. (\ref{nt}), (\ref{tau}) 
and relation $({\bf Pr})\dot{}={\bf P}^2=const$,
such a closed curve is smooth.},
the Stokes theorem gives
$
\oint_{\Gamma}\theta=\int_{S_+}d\theta=
-\int_{S_-}d\theta,
$
where $\Gamma=\partial S_+=-\partial S_-$,
$S_+\cup S_-=S^2$.
Within the path-integral quantization
scheme, the alternative representations 
for the same charge-monopole interaction term
in the action can differ only in $2\pi n,$ $n\in \ZZ$,
\beq
\label{nuq}
\int_{S_+}d\theta -\left(- \int_{S_-}d\theta
\right)=
\nu\int_{S^2}d\cos\vartheta\wedge d\varphi
=4\pi\nu=2\pi n,
\eeq
and we arrive at the Dirac quantization condition 
for the charge-monopole coupling constant:
$2\nu=n\in \ZZ$. 
Defining the dependent variables
$
s_i=-\nu n_i,
$
one can represent (\ref{ds2}) equivalently as
\beq
\label{omega}
\omega_{spin}=d\theta=-\frac{1}{2{\bf s}^2}\epsilon_{ijk}
s_ids_j \wedge ds_k,\quad
s_is_i=\nu^2.
\eeq
The two-form (\ref{omega}) is closed and nondegenerate,
and can be treated as a symplectic form
corresponding to the symplectic potential $\theta$.
If we drop out the kinetic term in 
the charge-monopole action (that corresponds to
taking the charge's zero
mass limit, $m\rightarrow 0$, \cite{dj}),
we get the CS action
\beq
\label{chsact}
S=\int_{t_1}^{t_2} \theta,
\eeq
describing the spin system. 
Indeed, for any function $f$ on the sphere $S^2$,
the symplectic form (\ref{omega}) defines 
the Hamiltonian vector field $X_f$ on the cotangent
bundle $T^*S^2$: 
$
i_{X_f}\omega(Y)=\omega(X_f,Y)=df.
$
This gives
$X_f=\partial_a f\omega^{ab}\frac{\partial}{\partial x^b}$,
and defines the corresponding Poisson brackets,
$
\{f,g\}=-\omega(X_f,X_g).
$
Here $x^a$, $a=1,2$, are the local coordinates on $S^2$ and
$\omega^{ab}$ are the elements of the matrix
inverse to the symplectic matrix $\omega_{ab}$, $\omega_{spin}=
\frac{1}{2}\omega_{ab}dx^a\wedge dx^b$.
Taking into account Eq. (\ref{omega}),
we get
$
\{s_i,s_k\}=\epsilon_{ijk}s_k,
$
$
s_is_i=\nu^2.
$
These relations define the classical spin system
with fixed spin modulus.
Geometric quantization \cite{gq1,gq5} applied to such
a system (for the details see ref. \cite{tor1})
leads to the same Dirac quantization of the parameter $\nu$,
$|\nu|=j$, $j=1/2,1,3/2,\ldots$,
and results in $(2j+1)$-dimensional 
representation of $su(2)$,
\beq
\label{su2p}
s_1=\frac{1-z^2}{2}\frac{d}{dz}+jz,\quad
s_2=i\frac{1+z^2}{2}\frac{d}{dz}-ijz,\quad
s_3=z\frac{d}{dz}-j,
\eeq
realized on the space of holomorphic 
functions with the basis 
$\psi_j^k\propto z^{j+k}$, $k=-j,-j+1,\ldots,+j$,
$s_3\psi_j^k=k\psi_j^k$,
and scalar product
\[
(\psi_1,\psi_2)=\frac{2j+1}{\pi}\int\int
\frac{\overline{\psi_1(z)}\psi_2(z)}{(1+
|z|^2)^{2j+2}}d^2z.
\]
The classical relation 
${\bf s}^2=\nu^2$ is changed for the quantum relation
${\bf s}^2=j(j+1)$.
Here the complex variable $z$ is related
to the spherical angles via the stereographic projection
$
z=\tan \frac{\vartheta}{2}e^{i\varphi}
$
from the north pole,
or via
$
z=\cot \frac{\theta}{2}e^{-i\varphi}
$
for the projection from the south pole.
In both cases
the symplectic two-form is represented as
\beq
\label{wzz}
\omega_{spin}=2i\nu\frac{d\bar{z}\wedge d{z}}{(1+\bar{z}z)^2}.
\eeq
Geometrically, the obtained spin system is
a  K\"ahler manifold \cite{Nakahara} with K\"ahler potential 
${\cal K}=2i\nu \ln(1+\bar{z}z)$:
$\omega_{spin}=\frac{\partial^2}{\partial z\partial\bar{z}}
{\cal K}\cdot d\bar{z}\wedge d{z}$, $\bar{z}=z^*$.
Locally, in  spherical coordinates
the spin Lagrangian is given by 
\beq
\label{lspinsing}
L_{spin}=\nu\cos\vartheta\dot{\varphi},
\eeq
and in terms of global complex variable
the Lagrangian (\ref{lspinsing}) takes the form
\beq
\label{lspin}
L_{spin}=i\nu \frac{\bar{z}\dot{z}-\dot{\bar{z}}z}{1+\bar{z}z}.
\eeq
The appearence of the two stereographic projections 
for the spin system (\ref{chsact}) reflects
the above mentioned necessity to work in two charts
in the case of the initial ($m\neq 0$) 
charge-monopole system to escape the
problems with Dirac string singularities.
In terms of globally defined independent variables $z$, $\bar{z}$
no gauge invariance left in the spin system
given by the Lagrangian (\ref{lspin})
but it is hidden in a fibre bundle structure
reflected, in particular, in the presence of two charts 
\cite{gq1}.

We conclude that the charge-monopole interaction term
has a nature of (0+1)-dimensional Abelian  CS term,
that leads to the
Dirac quantization of the charge-monopole coupling
constant, $|\nu|=j$.
In the limit of zero charge's mass
the total charge-monopole Lagrangian is reduced 
to the Lagrangian (\ref{lspin}) 
in terms of independent variables $z$ and $\bar{z}$,
which describes the spin-$j$ system. 
In what follows, we consider other possibilities 
to describe spin system proceeding from its nature
associated with the monopole CS term.

\section{Charge-monopole as a particle with spin}

The spin nature of the charge-monopole CS
term and observed free character of the charge's dynamics
allow us to get the alternative description
for the charge-monopole system as a 
free particle of fixed spin with translational
and spin degrees of freedom interacting via the helicity constraint.
To find such a description, we forget for the moment
that the spin system has been obtained via the identification
$
s_i=-\nu n_i,
$
(in the limit $m\rightarrow 0$),
and simply start with its Hamiltonian description
given by the symplectic form (\ref{omega}) 
and corresponding brackets 
$
\{s_i,s_k\}=\epsilon_{ijk}s_k.
$ 
The canonical Hamiltonian of the 
spin theory given by the first order action (\ref{chsact})
or by the Lagrangian (\ref{lspin}) 
is equal to zero.
Let us extend such a pure spin system
by adding to it independent translational 
degrees of freedom described by the particle's coordinates
$r_i$ and canonically conjugate momenta
$p_i$, i.e. we suppose 
that the corresponding symplectic two-form is
\beq
\label{omfree}
\omega=dp_i\wedge dr_i+\omega_{spin}.
\eeq
Moreover,  let  
the dynamics of the system is given by the Hamiltonian
$
H=\frac{1}{2}{\bf p}^2.
$
Then this Hamiltonian and 
relations (\ref{omega}), (\ref{omfree}) 
specify the nonrelativistic free particle 
with internal degrees of freedom describing spin of
fixed value.
In such a system we have $6+2$ independent
phase space variables instead of 6 variables 
in the initial charge-monopole system.
We can reduce such an extended system
to the initial system by introducing into it
one first class constraint.
Having in mind the identification 
$
s_i=-\nu n_i,
$
for the initial system (\ref{lchmo}),
let us postulate the helicity constraint
\beq
\label{heli}
\chi \equiv {\bf sr} + \nu r\approx 0.
\eeq
This constraint can be interpreted as a 
constraint introducing interaction between
translational and spin degrees of freedom.
But the constraint (\ref{heli}) is not conserved by
the Hamiltonian, $\frac{1}{2}\{{\bf p}^2,\chi\}\neq 0$,
and for consistency of such a theory
we have to  modify the latter.
For the purpose, let us find the complete set 
of gauge-invariant variables 
commuting in the sense of Poisson brackets 
with constraint (\ref{heli}).
Since we have $6+2$ phase space variables 
and one constraint, there are 6 independent 
gauge-invariant variables.
They are $r_i$ and 
\beq
\label{ginv}
\Pi_i\equiv p_i-\frac{1}{r^2}\epsilon_{ijk}r_j s_k,
\eeq
$\{r_i,\chi\}=0$, $\{\Pi_i,\chi\}\approx 0$.
Therefore, it is natural to change $H=\frac{1}{2}{\bf p}^2$ 
for its gauge-invariant analog,
$
H=\frac{1}{2}{\bf \Pi}^2,
$
and to take the sum of it and 
constraint (\ref{heli}) multiplied by an
arbitrary function $\lambda=\lambda(t)$, 
\beq
\label{ht}
H=\frac{1}{2}{\bf \Pi}^2 +\lambda \cdot ({\bf sr}+ \nu r),
\eeq
as a total Hamiltonian \cite{ht}.
The Poisson brackets 
for the gauge-invariant variables are
\beq
\label{poig}
\{\Pi_i,\Pi_j\}=-\frac{\bf rs}{r^4}\epsilon_{ijk}r_k,\quad
\{\Pi_i,r_j\}=\delta_{ij},\quad
\{r_i,r_j\}=0.
\eeq
Taking into account constraint (\ref{heli}),
the Poisson brackets between 
$\Pi_i$ and $\Pi_j$ coincide with the 
Poisson brackets (\ref{brmono}) between $P_i$ and $P_j$.
Identifying $\Pi_i$ with $P_i$, 
the dynamics generated by the Hamiltonian
(\ref{ht}) for the gauge-invariant variables $r_i$, $\Pi_i$
is exactly the same as the dynamics 
in the initial charge-monopole system (\ref{lchmo}),
and the physical content of the extended system
(\ref{ht}) is the same as that of the initial system
(\ref{lchmo}). In particular, the vector
\beq
\label{jtot}
{\bf J}={\bf r}\times {\bf p}+{\bf s}
\eeq
is the integral of motion, whose components
satisfy the Poisson bracket relations
$
\{
{J}_i,{J}_j\}=\epsilon_{ijk}{J}_k,
$
and generate rotations.
It can be represented equivalently
as
$
{\bf J}={\bf r}\times {\bf \Pi}+
({\bf ns})\cdot {\bf n},
$
and on the constraint surface
(\ref{heli}) takes the form
$
{\bf J}\approx {\bf r}\times{\bf \Pi}
-\nu \cdot {\bf n},
$
which coincides with (\ref{jint}) 
with identified $\Pi_i$ and $P_i$. 
With the help of relation (\ref{jtot}),
the Hamiltonian (\ref{ht})
can be written down equivalently 
(cf. with Eq. (\ref{hame3})),
\beq
\label{horb}
H=\frac{1}{2}p_r^2+\frac{({\bf J}\times {\bf n})^2}{2r^2}
+\lambda \cdot ({\bf J}{\bf n}+\nu),
\eeq
where $p_r={\bf pr}\cdot r^{-1}$ 
is the momentum canonically conjugate to the radial variable $r$.
Since ${\bf Ln}=0$, ${\bf L}={\bf r}\times {\bf p}$,
the difference
of the present Hamiltonian interpretation of the charge-monopole
system as a particle with spin 
from the reduced E(3) system from section 2.4 is
that here the helicity is fixed weakly by
the constraint (\ref{heli}), whereas there it was fixed 
strongly by
the second relation from Eq. (\ref{centr}).
As a consequence, here the components of the angular momentum vector
${\bf L}$ form the $so(3)$ algebra, 
$\{L_i,L_j\}=\epsilon_{ijk}L_k$,
but they, unlike the total angular momentum vector
${\bf J}$, are not physical variables:
$\{L_i,\chi\}\neq 0$. 
In the present interpretation, 
we have additional spin variables $s_i$
restricted by the condition ${\bf s}^2=\nu^2$,
but the only physical observable constructed 
from them is the combination 
${\bf sn}$, which is fixed by the helicity
constraint.
The important comment is in order here.
Strictly speaking,
in correspondence with the discussion above on the
necessity of restriction $J^2>\nu^2$,
we have to suppose that the relation ${\bf s}^2=\nu^2$
has to be treated only in the sense of the limit 
${\bf s}^2=\nu^2+\varepsilon^2$, $\varepsilon\rightarrow 0$.
This will not only correspond to the specified
restriction on $J^2$ in accordance with
relations (\ref{heli}) and (\ref{jtot}),
but is necessary for the consistent treatment of the
extended model as a constrained system.
Indeed, in correspondence with general theory 
of gauge systems \cite{ht},
only in this case the constraint (\ref{heli})
can be treated as a good constraint condition,
which in the two-dimensional spin phase subspace given by $s_i$,
$\{s_i,s_j\}=\epsilon_{ijk}s_k$,
${\bf s}^2=\nu^2 +\varepsilon^2$,
specifies one-dimensional physical subspace
(on which it will act transitively). 

The Lagrangian corresponding to the described extended
system is
\beq
\label{lfspin}
L=\frac{1}{2}\dot{\bf r}{}^2-
\frac{1}{r^2}({\bf r}\times \dot{\bf r})\cdot
{\bf s}
-\lambda \cdot ({\bf rs}+\nu r)
+L_{spin},
\eeq
with $L_{spin}$ chosen in the form (\ref{lspin}) 
and $\lambda$ treated as a Lagrange multiplier.

We conclude that the system
(\ref{lfspin}) describing the nonrelativistic
particle of spin $\nu$ with interacting translational and
spin degrees of freedom (see the second ${\bf Ls}$-coupling term
and the third constraint term in Lagrangian)
is classically equivalent to the charge-monopole
system (\ref{lchmo}).

The quantum theory of the charge-monopole system in such interpretation
is obvious. As we have seen, the  quantization of the
spin variables results in the Dirac condition,
$\nu=\epsilon j$, $\epsilon=+$ or $-$,
and gives rise to the corresponding $(2j+1)$-dimensional representation
of the $su(2)$ with spin operators (\ref{su2p}) acting on the
space of holomorphic functions.
Choosing the  representation diagonal in $r_i$ and
realizing $p_j$ as differential operators, 
$p_j=-i\partial/\partial r_j$,
one can work on the space of functions of the form
$\Psi^j({\bf r},z)=\sum_{k=-j}^{j}\psi_k({\bf r})z^{j+k}$.
The quantum analog of the classical constraint
takes the form of the quantum condition
separating the physical subspace,
$
({\bf sn}+\epsilon j)\Psi^j_{phys}({\bf r},z)=0.
$
It is necessary to note that quantum mechanically 
the above mentioned necessity of the regularization 
${\bf s}^2=\nu^2+\varepsilon^2$ is taken into account automatically.
Indeed, since the eigenvalues of the 
operators ${\bf s}^2$ and $({\bf sn})^2$ 
are separated in a necessary way, 
${\bf s}^2=j(j+1)$, $({\bf sn})^2=j^2$, 
one can say that the quantization
``cures" the classical system. 

\section{Charge-monopole system as a symmetric top}

In this section we show that the charge-monopole system
can also be interpreted as a reduced symmetric spinning top
with dynamical tensor of inertia and ``isospin" U(1) gauge symmetry.
To this end, we return to the spin symplectic 
form (\ref{ds2}), and supplement the unit vector ${\bf n}\equiv {\bf e}_3$
with the two vectors ${\bf e}_1$ and ${\bf e}_2$
forming together the oriented orthonormal set of vectors,
\beq
\label{basis}
{\bf e}_a{\bf e}_b=\delta_{ab},\quad
{\bf e}_1\times {\bf e}_2={\bf e}_3.
\eeq
As a consequence of basic relations (\ref{basis}),
the vectors ${\bf e}_a$, $a=1,2,3,$
satisfy also the completeness relation
\beq
\label{bascom}
e^i_ae^j_a=\delta^{ij}.
\eeq
Using Eqs. (\ref{basis}), (\ref{bascom}),
one can represent the two-form (\ref{ds2})
as
\beq
\label{dte}
d\theta=\nu d{\bf e}_1\wedge d{\bf e}_2,
\eeq
from which we get another representation
for the one-form,
\beq
\label{the}
\theta=\frac{\nu}{2}({\bf e}_1d{\bf e}_2-{\bf e}_2d{\bf e}_1).
\eeq
Taking into account the relation
$\dot{\bf e}_3^2=(\dot{\bf e}_1{\bf e}_3)^2+(\dot{\bf e}_2{\bf e}_3)^2$,
we get the alternative Lagrangian for the charge-monopole system,
\beqa
L&=&\frac{1}{2}\dot{r}{}^2+\frac{1}{2}r^2
\left(
(\dot{\bf e}_1\cdot{\bf e}_1\times{\bf e}_2)^2
+
(\dot{\bf e}_2\cdot{\bf e}_1\times{\bf e}_2)^2
\right)
+\frac{\nu}{2}({\bf e}_1\dot{\bf e}_2-
{\bf e}_2\dot{\bf e}_1)
\nonumber\\
&&-\frac{\lambda_1}{2}({\bf e}_1^2-1)
-\frac{\lambda_2}{2}({\bf e}_2^2-1)
-\lambda_{12}{\bf e}_1{\bf e}_2,
\label{ltop}
\eeqa
with $\lambda_1$, $\lambda_2$ and $\lambda_{12}$ being
Lagrange multipliers, variation over which
gives the Lagrangian constraints 
${\bf e}_1^2-1=0$, ${\bf e}_2^2-1=0$
and ${\bf e}_1{\bf e}_2=0$.
With these constraints, the equations of motion
for ${\bf r}= r{\bf e}_3$,
${\bf e}_3={\bf e}_1\times {\bf e}_2$,
following from (\ref{ltop}) coincide with the
charge-monopole Lagrange equations (\ref{chmr}).
To prove the complete equivalence
of the model (\ref{ltop}) 
to the initial system (\ref{lchmo}),
we pass over to the Hamiltonian formalism.
The canonical Hamiltonian corresponding to
Lagrangian (\ref{ltop}) is
\beq
\label{htopcan}
H_{c}=\frac{1}{2}p_r^2+\frac{1}{2r^2}({\bf J}\times{\bf e}_3)^2
+\frac{\lambda_1}{2}({\bf e}_1^2-1)+
\frac{\lambda_2}{2}({\bf e}_2^2-1)+\lambda_{12}{\bf e}_1{\bf e}_2,
\eeq
where ${\bf e}_3={\bf e}_1\times{\bf e}_2$
and the total angular momentum is 
${\bf J}={\bf e}_1\times{\bf p}_1+{\bf e}_2\times{\bf p}_2$
with ${\bf p}_{1,2}$ being the momenta canonically conjugate 
to ${\bf e}_{1,2}$. 
Note that in terms of the velocity phase space variables
the total angular momentum is reduced to
${\bf J}=r^2\cdot{\bf e}_3\times\dot{\bf e}_3-\nu{\bf e}_3$
in correspondence with Eqs. (\ref{jint}).
The application of Dirac-Bergmann theory to the system
(\ref{ltop}) results in the following complete
set of constraints:
\beqa
&\pi_1\approx 0,\quad
\pi_2\approx 0,\quad
\pi_{12}\approx 0,&\label{pi12}\\
&{\bf e}_1^2-1\approx 0,\quad
{\bf p}_1{\bf e}_1\approx 0,\quad
{\bf e}_2^2-1\approx 0,\quad
{\bf p}_2{\bf e}_2\approx 0,\quad
{\bf e}_1{\bf e}_2\approx 0,\quad
{\bf p}_1{\bf e}_2+{\bf p}_2{\bf e}_1\approx 0,&\label{ep12}\\
&{\bf p}_1{\bf e}_2-{\bf p}_2{\bf e}_1+\nu\approx 0,&\label{j3}
\eeqa
with the momenta $\pi_1$, $\pi_2$ and $\pi_{12}$ 
canonically conjugate to Lagrange multipliers.
Constraints (\ref{pi12}) mean that 
the Lagrange multipliers are pure gauge
variables and can be completely excluded,  e.g.,
by supplementing (\ref{pi12}) with the gauge conditions
$\lambda_1\approx0$, $\lambda_2\approx0$, $\lambda_{12}\approx0$.
The constraints 
(\ref{ep12}) form the subset of second class constraints,
whereas (\ref{j3}) is the first class constraint.
Reduction of the symplectic two-form of the system,
$\omega=d{\bf p}_1\wedge d{\bf e}_1+
d{\bf p}_2\wedge d{\bf e}_2+d p_r\wedge dr$,
to the surface of second class constraints 
(\ref{ep12})
is given by
\beq
\label{omred}
\omega=\frac{1}{2}d({\bf J}\times {\bf e}_a)\wedge
d{\bf e}_a+
dp_r\wedge dr,
\eeq
where the summation over $a=1,2,3$ 
is assumed.
The two-form (\ref{omred})
is the symplectic form on the reduced phase space
which is described by the basis of vectors ${\bf e}_a$,
$a=1,2,3,$ subject  to conditions (\ref{basis})
as strong relations,
by the angular momentum vector ${\bf J}$
and by the canonically conjugate radial variables $r$ and $p_r$.
{}From (\ref{omred})
we obtain the following Poisson-Dirac brackets
on the reduced phase space:
\beq
\label{brtop}
\{e^i_a,e_b^j\}=0,\quad
\{J^i,e^j_a\}=\epsilon^{ijk}e^k_a,\quad
\{ J^i,J^j\}=\epsilon^{ijk}J^k,\quad
\{r,p_r\}=1, 
\eeq
and all other brackets 
for radial variables $r$ and $p_r$ to be equal to zero.
The remaining constraint (\ref{j3})
takes the form
\beq
\label{iso3}
\chi={\bf Je}_3+\nu\approx 0,
\eeq
and the total Hamiltonian of the system 
is reduced to the canonical Hamiltonian 
extended by the first class constraint
multiplied by an arbitrary function $\lambda(t)$:
\beq
\label{htottop}
H=\frac{1}{2}p_r^2+\frac{1}{2r^2}({\bf J}
\times {\bf e}_3)^2 +\lambda \cdot({\bf Je}_3+\nu).
\eeq
To understand the physical sense of the obtained system,
let us define the scalar 
quantities 
$I_a\equiv 
-{\bf Je}_a$, $I_aI_a={\bf J}^2$.
They satisfy the following Poisson bracket relations:
\beq
\label{ii}
\{I_a,I_b\}=\epsilon_{abc}I_c,\quad
\{I_a,e^i_b\}=\epsilon_{abc}e^i_c,\quad
\{I_a,J^i\}=0.
\eeq
Since $I_a$ commute with $J_i$ and satisfy
$su(2)$ algebra, one can treat them as 
components of the isospin vector.
On the other hand, due to the basic relations (\ref{basis}),
the quantities $e^i_a$ can be treated 
as the elements of the group $SO(3)$.
Then the quantities $J^i$ have a sense of the
basis of the left-invariant vector fields on this group
whereas the quantities $-I_a$ can be identified with
the basis of the right-invariant vector fields \cite{Nakahara,razpl}.
Due to the relations (\ref{ii}),
the first class constraint (\ref{iso3}) generates
gauge transformations of the system 
which have a sense of SO(2)$\sim$U(1) isospin rotations
generated by $I_3$. Under such transformations
the vectors ${\bf J}$ and
${\bf e}_3$ are invariant.
The electromagnetic U(1) gauge invariance 
of the curvature form (\ref{curvat}) corresponding
to the CS charge-monopole coupling term
is changed here for the described SO(2)$\sim$U(1) 
gauge invariance
of the two-form (\ref{dte}) corresponding
to the CS form (\ref{the}). 
Since $({\bf J}\times{\bf e_3})^2=I_1^2+I_2^2$,
the system given by the Hamiltonian (\ref{htottop})
and brackets (\ref{brtop}) can be identified
as a symmetric spinning top with the symmetry
axis given by ${\bf e}_3$ and dynamical moment
of inertia ${\cal I}_1={\cal I}_2=r^2$.
The isospin U(1) gauge symmetry
generated  by the constraint (\ref{iso3})
means that the rotation about the symmetry
axis ${\bf e}_3$ 
is of pure gauge nature and, so, is not observable.

The formal difference of the symmetric spinning top system
from  the reduced E(3) system discussed in section 2.4
is that here the condition ${\bf Jn}={\bf Je}_3=-\nu$
appears in the form of first class constraint (weak
equality) unlike the strong equality in the case of 
the reduced E(3) system.
The physical content of both systems is exactly the same.
Indeed, besides the radial variables $r$ and $p_r$,
the only observable quantities (commuting in the sense 
of  Dirac-Poisson brackets with the constraint (\ref{iso3})) 
are the vector of angular momentum  ${\bf J}$
and the unit vector ${\bf e}_3\equiv {\bf n}$.
So, the E(3) system can also be treated as a symmetric
spinning top reduced by the action of the first
class constraint (\ref{iso3}).

The system is quantized as follows.
In correspondence with the classical relations (\ref{basis}), 
(\ref{bascom})
and the sense of the complete orthonormal set of vectors ${\bf e}_a$,
they can be parametrized by the three Euler angles $\alpha,\beta,\gamma$,
and we can choose a representation diagonal in these angle
variables and in the radial variable $r$.
Then the components of the operator ${\bf J}$
are realized in the form of linear differential
operators of angular variables \cite{top,top1}. 
An arbitrary state can be decomposed over the
complete basis of Wigner functions,
\[
\Psi(r,\alpha,\beta,\gamma)=\sum_{j,s,k}\psi_{j,s,k}(r)
D_{s,k}^j(\alpha,\beta,\gamma),
\]
where $ j$ takes either integer, $j=0,1,2,\ldots$,
or half-integer, $j=1/2,3/2,\ldots$, values,
and $s,k=-j,-j+1,\ldots,+j$,
${\bf J}^2D^j_{s,k}=j(j+1)$, $J_3D^j_{s,k}=sD^j_{s,k}$,
$I_3D^j_{s,k}=kD^j_{s,k}$.
The quantum analog of the first class constraint (\ref{iso3})
is transformed into the equation separating the physical states:
\beq
\label{qphys}
(I_3-\nu)\Psi_{phys}=0.
\eeq
This equation has nontrivial solutions only
for $\nu=n/2$, $n\in \ZZ$, i.e. we arrive once again at the
Dirac quantization condition,
and the corresponding solutions of Eq. (\ref{qphys})
have the form
\beq
\label{sphys}
\Psi_{phys}(r,\alpha,\beta,\gamma)=\sum_{j,s}{}'
\psi^j_{s,\frac{n}{2}}(r)D^j_{s,\frac{n}{2}}(\alpha,\beta,\gamma),
\eeq
where prime means that in the sum $j$ takes the values
$j=|n/2|,|n/2|+1,\ldots$. 
Once again, let us note that here the quantization
separates in the necessary way the value of the quantized parameter
$|\nu|=|\frac{n}{2}|$ and possible eigenvalues of the angular 
momentum operator in the physical subspace:
${\bf J}^2= j(j+1)>\nu^2$.

It is interesting to note that in the present
``spinning top" picture for the charge-monopole system
the {\it total} angular momentum is represented 
in terms of the isospin angular momentum as
${\bf J}=-{\bf e}_aI_a$,
whereas from the point of view of the field 
theoretical mechanism \cite{JackR,HastH},
only {\it spin} part of the total angular
momentum vector of the charge-monopole system is 
created from the isospin degrees
of freedom. The seeming contradiction is explained
by the constraint (\ref{qphys}) prescribing
the total angular momentum to take the values
starting from the minimal value $|n/2|=|\nu|$,
which can be interpreted as the ``internal"
spin of the charge-monopole system,
and in this sense here spin is also created by isospin.

\section{Higher-derivative form of CS term}

The charge-monopole system
can also be described by the Lagrangian
with CS term represented in a higher-derivative form.
To show this, we write down the  CS form (\ref{the}) 
as $\theta=-\nu {\bf e}_2 d{\bf e}_1$.
Identifying the vectors 
${\bf e}_3$ and
${\bf e}_1$ 
with the unit vectors 
${\bf n}={\bf r}\cdot r^{-1}$ and
$\dot{\bf n}\cdot \vert \dot{\bf n}\vert{}^{-1}$,
respectively,
and taking into account that 
${\bf e}_2={\bf e}_3\times {\bf e}_1$,
the charge-monopole coupling term can be written down
in the equivalent higher-derivative form
\beq
\label{nnn}
L_{int}=-\frac{\nu}{\dot{\bf n}^2}{\bf n}\cdot (\dot
{\bf n}\times
\ddot{\bf n}).
\eeq
Simple geometrical consideration shows that the higher derivative
term $({\bf n}\times\dot{\bf n})\cdot\ddot{\bf n}/\dot{\bf n}{}^2$
has a sense of angular velocity of rotation of the vector
$\dot{\bf n}$ about the vector ${\bf n}$. 
The total charge-monopole Lagrangian is given by
\beq
\label{lhigh}
L=\frac{1}{2}\dot{\bf r}{}^2-\nu
\frac{r}{({\bf r}\times \dot{\bf r})^2}({\bf r}\times\dot
{\bf r})
\cdot\ddot{\bf r}.
\eeq
Naively, the equations of motion following from
Lagrangian (\ref{lhigh}), 
\beq
\label{hieq}
\frac{\partial L}{\partial {\bf r}}-
\frac{d}{dt}\frac{\partial L}{\partial \dot{\bf r}}+
\frac{d^2}{dt^2}\frac{\partial L}{\partial\ddot{\bf r}}=0,
\eeq
are the third order differential equations
and their equivalence to the second-order
equations (\ref{chmr}) is not  obvious.
Let us prove  the equivalence of equations of motion 
(\ref{hieq}) to Eq. (\ref{chmr}).
First we note that the higher-derivative term
(\ref{nnn}) satisfies the following 
relations:
\beqa
&{\bf r}\cdot
{\partial L_{int}}/{\partial{\bf r}}=0,\quad
\dot{\bf r}\cdot
{\partial L_{int}}/{\partial\dot{\bf r}}=-L_{int},\quad
\ddot{\bf r}\cdot
{\partial L_{int}}/{\partial\ddot{\bf r}}=+L_{int},&
\nonumber\\
&
\dot{\bf r}\cdot
{\partial L_{int}}/{\partial{\bf r}}=
({{\bf r}\dot{\bf r}}){r}^{-2}\cdot L_{int},\quad
{\bf r}\cdot
{\partial L_{int}}/{\partial\dot{\bf r}}=
{\bf r}\cdot
{\partial L_{int}}/{\partial\ddot{\bf r}}=
\dot{\bf r}\cdot
{\partial L_{int}}/{\ddot{\bf r}}=0.&
\label{proint}
\eeqa
Multiplying the equations (\ref{hieq})
subsequently by the vectors ${\bf r}$,
$\dot{\bf r}$ and ${\bf r}\times\dot{\bf r}$, 
and using Eqs. (\ref{proint}),
we arrive at the three equations
$
\ddot{\bf r}{\bf r}=0,
$
$
\ddot{\bf r}\dot{\bf r}=0,
$
$
\ddot{\bf r}\cdot({\bf r}\times\dot{\bf r})=-\frac{\nu}{r^3}
({\bf r}\times\dot{\bf r})^2.
$
Since ${\bf r}$, 
$\dot{\bf r}-{\bf r}({\bf r}\dot{\bf r})\cdot r^{-2}$
and ${\bf r}\times \dot{\bf r}$ form the complete basis of orthogonal
vectors, we conclude that the equations of motion 
(\ref{hieq}) are equivalent to the obtained system of three scalar equations, 
which, in turn, is equivalent to one vector equation (\ref{chmr}).
Therefore, the coupling of 
the charge to the magnetic monopole can be alternatively 
described by the scale invariant higher derivative term (\ref{nnn}).

It is worth noting that analogously to the present system,
the relativistic massive particle in (2+1) dimensions
coupled to the external constant homogeneous
electromagnetic field turns out to be classically equivalent \cite{tor4}
to the higher derivative model of relativistic
particle with torsion given by the action \cite{Pol,tor1}
$
S_{tor}=-\int(m+\alpha\kappa)ds,
$
where $ds^2=-dx_\mu dx^\mu$, $\kappa$ is a scale invariant
torsion of the particle's world trajectory,
$\kappa=\epsilon^{\mu\nu\lambda}x'_\mu x''_\nu x'''_\lambda/(x'')^2$,
$x'_\mu=dx_\mu/ds$, and $\alpha$ is a parameter.
Such a model underlies (2+1)-dimensional anyons \cite{any0,tor1},
and in the last section we shall discuss 
the relationship between the charge-monopole system
and anyons.

\section{Twistor description of the charge-monopole system}

The helicity constraint appearing in the charge-monopole
system interpreted as a particle with spin 
is analogous to the helicity-fixing constraint 
in the case of massless particle with spin.
The latter system admits, in particular,
the twistor description \cite{t1}-\cite{t4}.
Using this observation, one can get
the twistor formulation for the charge-monopole system.

In the twistor approach for the massless particle,
the corresponding energy-momentum vector is treated as a 
``composite" object constructed from the
twistor (even spinor) variables, 
and the helicity fixing constraint 
generates the U(1) transformations for twistors.
By analogy, let us introduce mutually
conjugate even complex variables $z_a$ and $\bar{z}_a=z_a^*$,
$a=1,2$, forming two conjugate spinors. With them, we
represent the 
charge coordinate vector ${\bf r}$ as a composite
vector,
\beq
\label{rzz}
\varphi_i\equiv r_i-z\sigma_i\bar{z}\approx 0,
\eeq
where $\sigma_i$ is the set of Pauli matrices.
Introducing the momenta ${\cal P}_a$, $\bar{\cal P}_a$
canonically conjugate to the spinor variables,
$\{z_a,{\cal P}_b\}=\delta_{ab}$,
$\{\bar{z}_a,\bar{\cal P}_b\}=\delta_{ab}$,
we construct the generators of rotations (the total angular momentum vector),
\beq
\label{jtw}
{J}_i=({\bf r}\times{\bf p})_i+\frac{i}{2}\left(
z{\sigma}_i{\cal P}-\bar{\cal P}{\sigma}_i\bar{z}\right),
\eeq
forming the su(2) algebra: 
$\{J_i,J_j\}=\epsilon_{ijk}J_k$.
Here ${\bf p}$ is the vector canonically conjugate
to ${\bf r}$. 
Following the twistor approach, 
we also introduce 
the constraint
\beq
\label{zpa}
\chi\equiv \frac{i}{2}(\bar{z}\bar{\cal P}-z{\cal P})-\nu\approx 0,
\eeq
which is in involution with the constraints (\ref{rzz}) and
generates the U(1) gauge transformations for the spinor variables,
$z_a\rightarrow z'_a=e^{i\gamma}z_a$,
$\bar{z}_a\rightarrow \bar{z}{}'_a=e^{-i\gamma}\bar{z}_a$,
${\cal P}_a\rightarrow {\cal P}'_a=e^{-i\gamma}{\cal P}_a$,
$\bar{\cal P}_a\rightarrow \bar{\cal P}{}'_a=e^{i\gamma}\bar{\cal P}_a$,
with $\gamma=\gamma(t)$ being a parameter of transformation.
Taking into account the identity
\beq
\label{ss}
\sigma^i_{ab}\sigma^i_{cd}=\delta_{ab}\delta_{cd}-
2\epsilon_{ac}\epsilon_{bd},
\eeq
with $\epsilon_{ab}=-\epsilon_{ba}$, $\epsilon_{12}=1$,
we get the relation $r\approx\bar{z}z$ 
as the consequence of the constraint (\ref{rzz}).
This relation and Eq. (\ref{jtw}) allow us to 
represent the constraint (\ref{zpa}) in the equivalent form
of the helicity-fixing constraint, 
\beq
\label{tchi}
\tilde{\chi}\equiv {\bf Jr}+\nu r\approx 0.
\eeq
Since the total phase space described by $r_i$, $p_i$,
$z_a$, $\bar{z}_a$, ${\cal P}_a$ and $\bar{\cal P}_a$
is $14$-dimensional, and we have 4 first class constraints 
(\ref{rzz}) and (\ref{zpa}),
there are only 6 independent physical phase space degrees of freedom
like in the charge-monopole system.
They are given by the observables having zero Poisson brackets
with all the constraints.
Such independent variables are $r_i$ and 
\beq
\label{ptwis}
\Pi_i\equiv
p_i+\frac{1}{2\bar{z}z}(z\sigma_i{\cal P}+\bar{\cal P}\sigma_i\bar{z}),
\eeq
for which we have the Poisson bracket relations $\{r_i,r_j\}=0$,
$\{r_i,\Pi_j\}=\delta_{ij}$ and
\beq
\label{pptw}
\{\Pi_i,\Pi_j\}=(\bar{z}z)^{-2}\epsilon_{ijk}
({\bf r}\times{\bf \Pi}-{\bf J})_k\approx\frac{\nu}{r^3}\epsilon_{ijk}r_k.
\eeq
The weak equality means the equality on the surface
of constraints (\ref{rzz}) and (\ref{zpa}).
Physical variables $\Pi_i$ 
correspond here to the charge-monopole variables $P_i$.
The Hamiltonian of the charge-monopole
system in the twistor formulation 
can be taken as the linear combination of ${\bf \Pi}^2$
and of the first class constraints,
\beq
\label{htwi}
H=\frac{1}{2}{\bf \Pi}^2+\rho_i\cdot \varphi_i+\lambda\cdot \chi,
\eeq
with $\rho_i=\rho_i(t)$, 
$\lambda=\lambda(t)$ being arbitrary functions
(Lagrange multipliers).
Direct calculation shows that 
the equations of motion generated by this Hamiltonian
are reduced to $\ddot{\bf r}\approx -\nu r^{-3}\cdot{\bf r}\times
\dot{\bf r}$, i.e. the Hamiltonian
(\ref{htwi}) and constraints
(\ref{rzz}), (\ref{zpa}) give the alternative
twistor description for the charge-monopole system.

The quantum theory of the system in the twistor 
approach is the following.
It is natural to choose the represenation diagonal
in ${\bf r}$,  $z_a$ and $\bar{z}_a$, and realize the 
canonically conjugate momenta in the form of differential operators.
In accordance with constraint (\ref{rzz}),
the physical states have to be of the form
$\Psi_{phys}=\delta^{(3)}(\bsymb{r}-z\bsymb{\sigma}\bar{z})\cdot
\psi(z,\bar{z})$. 
Then like in the model of massless particle with spin \cite{t2},
the quantum analog of the constraint (\ref{zpa}),
\beq
\label{twsp}
\left(\frac{1}{2}\left(\bar{z}_a\frac{\partial}{\partial \bar{z}_a}
-z_a\frac{\partial}{\partial z_a}\right)-\nu 
\right)\Psi_{phys}=0,
\eeq
and the requirement of single-valuedness of the wave functions
results in the  quantization of the charge-monopole constant,
$2\nu=n$, $n\in \ZZ$.
Finally, the physical states subject to Eq. (\ref{twsp})
will be described by the
wave functions of the form
\beq
\label{twpsi}
\Psi_{phys}=\delta^{(3)}(\bsymb{r}-z\bsymb{\sigma}\bar{z})
\cdot\psi_{phys}(z,\bar{z}),\quad
\psi_{phys}(z,\bar{z})=
\sum_{k=-\infty}^\infty \sum_{a,b=1,2}
C_k^{ab}\cdot (z_a)^k(\bar{z}_b)^{n+k},
\eeq
where $C^{ab}_k$ are constants.
The action of quantum analogs of classical observables 
${\bf r}$ and ${\bf \Pi}$ on physical wave functions 
(\ref{twpsi}) is reduced to 
\beqa
{\bf r}\Psi_{phys}&=&\delta^{(3)}(\bsymb{r}-z\bsymb{\sigma}\bar{z})
\cdot z\bsymb{\sigma}\bar{z}\cdot\psi_{phys}(z,\bar{z}),\nonumber\\
{\bf \Pi}\Psi_{phys}&=&
-i\delta^{(3)}(\bsymb{r}-z\bsymb{\sigma}\bar{z})
\cdot (2\bar{z}z)^{-1}(z\bsymb{\sigma}\frac{\partial}{\partial z}+
\frac{\partial}{\partial\bar{z}}\bsymb{\sigma}
\bar{z})\cdot\psi_{phys}(z,\bar{z}).
\label{rpred}
\eeqa

By inverse Legendre transformation, analogously to the 
case of the massless particle with spin \cite{t2},
one can construct the Lagrangian corresponding to Hamiltonian (\ref{htwi}).
Instead of realizing such constructions, we note here that
the list of independent  observable quantities and the action
of their quantum analogs on the physical states 
indicate  on the possibility to exclude 
${\bf r}$ and ${\bf p}$ as independent variables
and to construct the theory realized
only in terms of spinor variables.
To get the corresponding twistor form of the Lagrangian,
it is convenient to start from the higher-derivative 
Lagrangian (\ref{lhigh}) by putting in it 
$
\bsymb{r}=z\bsymb{\sigma}\bar{z}.
$
The remarkable feature of such a twistor representation
is that its substitution into the 
CS term of the higher derivative form (\ref{nnn})
reduces the latter to 
$-i\nu(\dot{\bar z}z-\bar{z}\dot{z})/\bar{z}z$,
and we arrive at the Lagrangian
\beq
\label{lzz}
L=\frac{1}{2}\left(\frac{d}{dt}(z\bsymb{\sigma}\bar{z})\right)^2+i\nu
\frac{\bar{z}\dot{z}-\dot{\bar{z}}z}{\bar{z}z}.
\eeq
Let us show that (\ref{lzz}) 
describes the charge-monopole system.
{}From the definition of the canonical momenta
${\cal P}_a=\partial L/\partial \dot{z}_a$,
$\bar{\cal P}_a=\partial L/\partial \dot{\bar{z}}_a$,
we find that the  constraint (\ref{zpa}) is the primary
constraint for the system (\ref{lzz}) and the 
total Hamiltonian is 
\beq
\label{hzz}
H=\frac{1}{2z\bar{z}}
\pi\bar{\pi}
+\lambda\cdot \chi,
\eeq
where we have introduced the notation
$
\pi_a={\cal P}_a-i\nu\bar{z}_a\cdot(\bar{z}z)^{-1},
$
$
\bar{\pi}_a=\bar{\cal P}_a+i\nu z_a\cdot(\bar{z}z)^{-1}.
$
The constraint (\ref{zpa})
is conserved by the Hamiltonian and, as a consequence, there are no
secondary constraints.
Therefore, the constraint (\ref{zpa}) is the first class constraint
and the number of physical phase space degrees of freedom
is equal to $8-2=6$.
The corresponding 6 independent observables weakly commuting in 
the Poisson bracket sense with the constraint (\ref{zpa})
are 
\beq
\label{piptwi}
r_i= z\sigma_i\bar{z},\quad 
\Pi_i=\frac{1}{2z\bar{z}}(\bar{\cal P}\sigma_i\bar{z}+
z\sigma_i{\cal P}).
\eeq
These quantities satisfy the following Poisson bracket relations:
$\{r_i,r_j\}=0$, $\{r_i,\Pi_j\}=\delta_{ij}$,
$\{\Pi_i,\Pi_j\}\approx\nu r^{-3}\epsilon_{ijk}r_k$,
where the weak equality means the equality on the surface
of the constraint (\ref{zpa}).
Obviously, they have the sense of the charge-monopole variables
$r_i$ and $P_i$ represented in the composite
form (\ref{piptwi}) in terms of twistor variables.
The variables $\Pi_i$ are
nothing else as the classical analogs of the 
quantum operators corresponding to 
observables (\ref{ptwis}), whose action is reduced to the surface
of the constraints (\ref{rzz}), i.e. they are
the classical analogs of operators (\ref{rpred}).
The conserving $su(2)$ generators are represented here
in the form 
$
{\bf J}={\bf r}\times{\bf \Pi}-\nu{\bf n}
$
with ${\bf n}={\bf r}\cdot r^{-1}$ and 
$r_i$, $\Pi_i$  given by Eq. (\ref{piptwi}).
Direct verification shows that the Hamiltonian
(\ref{hzz}) generates the correct equations of motion,
$\ddot{\bf r}=-\nu r^{-3}{\bf r}\times\dot{\bf r}$.
It is worth noting that
the system (\ref{hzz}) can be treated as a
reduction of the Hamiltonian system (\ref{htwi}) \cite{ht,razpl}
to the surface of the constraints (\ref{rzz})
supplied with the gauge conditions 
$p_i-\frac{1}{2}(z\bar{z})^{-1}(z\sigma_i{\cal P}
+\bar{\cal P}\sigma_i\bar{z})\approx 0$.

To conclude the discussion of the twistor formulation
for the charge-monopole system, let us show how
the limit of zero mass described here by 
the Lagrangian
\beq
\label{twispil}
L=i\nu
\frac{\bar{z}\dot{z}-\dot{\bar{z}}z}{\bar{z}z}
\eeq
gives rise to the pure spin system.
Here, as before, it is assumed that the 
initial configuration space is $\CC^2-\{0\}$.
This system can be quantized by the method  
advocated in ref. \cite{FJ},
but the symplectic two-form corresponding to Lagrangian
(\ref{twispil}) is singular, 
$\omega=\omega_{ab}d\bar{z}_a\wedge dz_b$,
$\omega_{ab}=2i\nu(\bar{z}z)^{-1}\cdot
(\delta_{ab}-z_a\bar{z}_b\cdot(\bar{z}z)^{-1})$,
$\omega_{ab}\bar{z}_a=\omega_{ab}z_b=0$,
that complicates the analysis.
There is a more short way to reveal a spin nature of
the system (\ref{twispil}) by showing its equivalence
to the system (\ref{lspin}).
For the purpose we note that
in the regions where either $z_1\neq 0$ (chart $U_1$)
or $z_2\neq 0$ (chart $U_2$), one can define the 
complex coordinate $Z$ as
$Z=z_2/z_1$ or $Z=z_1/z_2$, respectively.
Then Lagrangian (\ref{twispil}) is represented equivalently as
$
L=i\nu(\bar{Z}\dot{Z}-\dot{\bar{Z}}Z)\cdot(1+\bar{Z}Z)^{-1}
-2\nu\dot{\varphi},
$
where $\varphi$ is the phase of the coordinate 
$z_a$ in the chart $U_a$, $a=1,2$.
The first term coincides exactly with the Lagrangian
(\ref{lspin}) corresponding to the pure spin system.
The second total derivative term is not important classically as well as
quantum mechanically since its contribution
to the action, $\Delta S=-4\nu\pi$, for periodical trajectories
is trivialized, $\Delta S=0(mod\, 2\pi)$,
if we take into account the quantization condition $2\nu\in\ZZ$.
Therefore, the system (\ref{twispil}) 
is equivalent to the pure spin system (\ref{lspin}).

The equivalence of the system (\ref{twispil}) to the system
(\ref{lspin}) is encoded in its gauge symmetries. 
In correspondence with the above mentioned degeneracy 
of the symplectic form, the system (\ref{twispil})
possesses two gauge invariances:
its action is invariant under the transformations 
$z_a\rightarrow \rho z_a$,
$\rho=\rho(t)>0$, 
generated by the constraint
$\rho=\bar{\cal P}\bar{z}+z{\cal P}\approx 0$,
and $z_a\rightarrow e^{i\varphi}z_a$,
$\varphi=\varphi(t)\in \RR$,
generated by the helicity constraint of the form (\ref{zpa}).
This means that the points $z_a$ and $\zeta z_a$,
$\zeta\in\CC,$ $\zeta\neq 0$,  in configuration space are
physically equivalent and should be identified, $z_a\sim\zeta z_a$.
As a result, the configuration space of the system
is the projective complex plane $\CC P^1=(\CC^2-\{0\})/\sim$,
on which the coordinate $Z$ introduced above plays
a role of the ``inhomogeneous" coordinate \cite{Nakahara}.

\section{Charge-monopole in a spherical geometry and spin} 
Let us restrict the motion of the charge in the monopole field
to the sphere ${\bf r}^2=1$.
This can be done by treating the relation 
$
{\bf r}^2-1=0
$
as a Lagrangian constraint,
$
L=\frac{1}{2}\dot{\bf r}{}^2+e\dot{\bf r}{\bf A}({\bf r})-
\frac{\lambda}{2}({\bf r}^2-1),
$
whose condition of conservation 
generates the Hamiltonian constraint $p_r=0$. 
The reduction to the surface
of these second class constraints excludes 
the radial variables $r$
and $p_r$, and we arrive at the Hamiltonian
describing the charge on the sphere with the monopole in its 
center\footnote{See also refs. \cite{sp1,sp2}, where some aspects of 
the charge-monopole system in a spherical geometry were discussed.},
$
H=\frac{1}{2}({\bf J}\times{\bf n})^2.
$
Here the dynamical variables ${\bf J}$ and ${\bf n}$
satisfy the Poisson bracket relations (\ref{njPoi})
and are subject to the conditions (\ref{centr}),
i.e. the reduced system is a pure reduced E(3) system.
The same procedure of reduction 
can be realized in a spinning top picture,
i.e. by adding the condition ${\bf r}^2-1=0$  to 
Lagrangian (\ref{ltop}) as a Lagrangian constraint,
or by putting $r=1$ directly in Lagrangian (\ref{ltop}):
\beq
\label{topr1}
L=\frac{1}{2}\dot{\bf n}{}^2+\frac{\nu}{2}
({\bf e}_1\dot{\bf e}_2-{\bf e}_2\dot{\bf e}_1),
\eeq
where we suppose that ${\bf n}={\bf e}_1\times{\bf e}_2$
and vectors ${\bf e}_1$ and ${\bf e}_2$ are orthonormal.
In the case of higher derivative treatment of the charge-monopole 
system the corresponding reduced Lagrangian
is
\beq
\label{sfhi}
L=\frac{1}{2}\dot{\bf n}{}^2-\frac{\nu}{\dot{\bf n}{}^2}
{\bf n}\cdot(\dot{\bf n}\times\ddot{\bf n}),
\eeq
whereas in the twistor picture
the Lagrangian takes the form
\beq
\label{sferaz}
L=-2(\dot{z}\sigma_2 z)\cdot 
(\dot{\bar z}\sigma_2 \bar{z})+
i\nu (\bar{z}\dot{z}-\dot{\bar z}z),\quad
\bar{z}z=1.
\eeq
The condition $\bar{z}z=1$ can be omitted
by normalizing appropriately the Lagrangian,
i.e. by multiplying the first and second terms 
by $(\bar{z}z)^{-2}$ and 
$(\bar{z}z)^{-1}$, respectively.

The CS term 
is not changed by the reduction procedure
due to its scale-invariance.
But this term in addition is reparametrization invariant,
whereas the total Lagrangian has no reparametrization 
invariance.
We can  change the non-invariant 
second order in velocity term $\frac{1}{2}\dot{\bf n}{}^2$
for the first order term $\sqrt{\dot{\bf n}{}^2}$, that
gives rise to the reparametrization invariant action.
Let us analyze the physical content of such reparametrization action
considering, e.g., the modification of
the Lagrangian (\ref{topr1}),
\beq
\label{spinrep}
L=\gamma \sqrt{\dot{\bf n}{}^2}+
\frac{\nu}{2}({\bf e}_1\dot{\bf e}_2-{\bf e}_2\dot{\bf e}_1),
\eeq
where $\gamma>0$ is a dimensionless parameter.
{}The canonical Hamiltonian of the system (\ref{spinrep})
is equal to zero, and from the Hamiltonian point of view, 
the difference of the system (\ref{spinrep})
from the system (\ref{topr1}) consists in the presence 
of the constraint
$
{\bf J}^2-\kappa^2\approx 0,
$
$
\kappa^2\equiv\gamma^2+\nu^2,
$
generating reparametrizations
in addition to the isospin U(1) gauge symmetry
generated by the constraint
(\ref{iso3}).
The system (\ref{spinrep})
has the same physical content as 
the spin system of section 3 given by only
the CS term.
Indeed, applying the quantization scheme of section 5,
the physical subspace is separated here by 
Eq. (\ref{qphys})
and by
$
({\bf J}^2-\kappa^2)\Psi_{phys}(\alpha,\beta,\gamma)=0.
$
These two equations have 
nontrivial solutions only when the parameters $\kappa$ and
$\nu$ are quantized:
$\kappa^2=j(j+1)$ and $\nu=k$, where $k$, $|k|\leq j$, is integer 
(half-integer) for $j$ integer (half-integer),
and the corresponding physical state is 
$
\Psi_{phys}(\alpha,\beta,\gamma)=
\sum_{s=-j}^jC_s D_{s,k}^j(\alpha,\beta,\gamma),
$
where $C_s$ are constants.
This wave function describes an 
arbitrary state of fixed spin $j$ in the form 
alternative to the holomorphic functions of section 3.
Therefore, the reparametrization-invariant
system (\ref{spinrep}) is equivalent to the 
spin system described by only the CS term.

The charge-monopole 
system in spherical geometry can be treated
as a partially gauge fixed version of the 
reparametrization and scale invariant spin system (\ref{spinrep}).
To see this, we rewrite the Lagrangian
(\ref{spinrep}) in equivalent form by
introducing the einbein $v$:
\beq
\label{ein}
L=\frac{\dot{\bf n}{}^2}{2v}+\frac{v}{2}\gamma^2 
+
\frac{\nu}{2}({\bf e}_1\dot{\bf e}_2-{\bf e}_2\dot{\bf e}_1).
\eeq
Reparametrization invariance of the system (\ref{ein})
can be fixed (locally, see ref. \cite{razpl})
via introducing the appropriate gauge-fixing
conditions for the constraint 
$
{\bf J}^2-\kappa^2\approx 0,
$
and for the constraint $p_v\approx 0$,
where $p_v$ is the momentum canonically conjugate to
$v$.  On the other hand, introducing only the condition
$v=1$, we obtain the 
partially gauge fixed version of the system (\ref{ein})
\cite{ht}.
The Lagrangian (\ref{ein}) with $v=1$
is the Lagrangian (\ref{topr1}) 
shifted for the inessential constant.
Therefore, the charge-monopole system in a spherical
geometry can formally be treated  as a partially gauge fixed 
version of the spin system (\ref{ein}).

Analogously, we can change the 
kinetic term $\frac{1}{2}\dot{\bf r}^2$
in the initial charge-monopole
Lagrangian (\ref{lchmo}) for the 
reparametrization invariant term
$\sqrt{\dot{\bf r}{}^2}$, the latter is a 
kinetic term of relativistic particle
in 3D Euclidean space. As a result, 
we get the reparametrization
invariant action
\beq
\label{repaE}
S=\int L_{r}dt,\quad 
L_{r}=\frac{\dot{\bf r}{}^2}{2v}+\frac{v}{2} 
+e{\bf A}\dot{\bf r}.
\eeq
Its  partial gauge fixed version ($v=1$)
will give the Lagrangian coinciding up to inessential constant
with the initial Lagrangian (\ref{lchmo}).
{}From this point of view, the time translation,
the time dilation and special conformal 
transformation symmetries produced canonically by the $so(2,1)$
generators $H$, $D$ and ${\cal R}$ \cite{so2,jackvort}, 
can be treated as a relic of the reparametrization symmetry
of the system (\ref{repaE})
surviving the 
described formal Lagrangian gauge fixing procedure.

\section{Charge-monopole system and anyons}

Let us discuss the relationship
between the charge-monopole system and (2+1)-dimensional
anyons in the light of the obtained results.
Earlier, the analogy with the charge-monopole system
played an important role 
in constructing the theory of anyons as spinning particles 
\cite{any3}-\cite{tor5}.

We have observed that the charge-monopole system 
in many aspects is similar to the 3D free particle.
In (2+1)-dimensions, spin is a (pseudo)scalar and, 
as a consequence, the anyon of fixed spin has the same number
of degrees of freedom as a spinless free massive particle.
The relationship between the 
charge-monopole and anyon can be understood better
within the framework of the canonical description of these two systems.
As we have seen, the charge-monopole system essentially 
is a reduced E(3) system. In the case of anyons,
the E(3) group is changed for the Poincar\'e group ISO(2,1). 
The translation generators of the corresponding groups
are ${\bf n}$  and $p_\mu$, the latter being the 
energy-momentum vector of the anyon, and the
corresponding Casimir central elements are fixed
by the relations ${\bf n}^2=1$ and 
\beq
\label{mshell}
p^2+m^2\approx 0,
\eeq
where $m$ is a mass of the anyon.
The rotation (Lorentz) generators are given by ${\bf J}$ and by
\beq
\label{jxp}
{\cal J}_\mu=\epsilon_{\mu\nu\lambda}x^\nu p^\lambda
+J_\mu, 
\eeq
where $J_\mu$  are the translation invariant
$so(2,1)$ generators satisfying the algebra of the form
(\ref{so21}), 
$\{J_\mu,J_\nu\}=\epsilon_{\mu\nu\lambda}J^\lambda$, 
and subject to the relation
$J_\mu J^\mu=-\alpha^2=const$. 
The representation (\ref{jxp}) for the 
anyon total angular momentum vector is,
obviously, the analog of the relation ${\bf J}={\bf L}+{\bf s}$
appearing under interpretation of the charge-monopole
system as a particle with spin.
In the charge-monopole system, the second Casimir element
of E(3) is fixed either strongly, ${\bf Jn}=-\nu$,
or in the form of the weak relation 
${\bf Jn}+\nu\approx 0$ (see Eqs. (\ref{heli}), (\ref{jtot})).
In the anyon model, spin also can be fixed either
strongly, ${\cal J}p=Jp=-\alpha m$, or in the form of the
weak (constraint) relation 
\beq
\label{spinany}
\chi_a\equiv Jp+\alpha m\approx 0.
\eeq
When the helicity is fixed strongly,
for the charge-monopole system the
symplectic form corresponding to the Poisson brackets
(\ref{brmono}), 
has a nontrivial contribution describing
noncommuting quantities $P_i$ being 
the components of the charge's velocity: 
$\omega=dP_i\wedge dr_i+\frac{\nu}{2r^3}\epsilon_{ijk}r_idr_j\wedge
dr_k$.
In the anyon case, the strong spin fixing
gives rise to the nontrivial Poisson structure
for the particle's coordinate's \cite{any3}-\cite{tor5}, 
\beq
\label{anyx}
\{x_\mu,x_\nu\}=
\alpha (-p^2)^{-3/2}\cdot \epsilon_{\mu\nu\lambda}p^\lambda.
\eeq
On the other hand, when we treat the charge-monopole
system as a particle with spin (helicity is fixed weakly),
one can work in terms of the canonical symplectic
structure for the charge's coordinates and momenta
(see Eq. (\ref{omfree})), but the canonical momenta
${\bf p}$ are not observables due to their
non-commutativity with the helicity constraint,
whereas the gauge-invariant extension of ${\bf p}$
given by Eq. (\ref{ginv}) plays the role of
the non-commuting quantities $P_i$.
Exactly the same picture takes place in
the case of anyon when its spin is fixed
weakly: the coordinates $x_\mu$ commute in this case,
$\{x_\mu,x_\nu\}=0$,
but they have nontrivial Poisson brackets 
with the spin constraint (\ref{spinany}),
whereas their gauge-invariant extension,
$X_\mu=x_\mu+\frac{1}{p^2}\epsilon_{\mu\nu\lambda}p^\nu J^\lambda$
(cf. with Eq. (\ref{ginv})), $\{X_\mu,\chi_a\}\approx 0$,
are non-commuting and 
reproduce the Poisson bracket relation (\ref{anyx}).
Like in the anyon case, the advantage
of the extended formulation for the 
charge-monopole system (when we treat it as a particle
with spin), is in the existence of canonical
charge's coordinates $r_i$ and momenta $p_i$.
Within the initial minimal formulation 
(given in terms of $r_i$ and gauge-invariant 
variables $P_i$), the canonical momenta $p_i$
are reconstructed from $P_i$ only locally,
$p_i=P_i+eA_i$, due to the global 
Dirac string singularities hidden in 
the monopole vector potential.
Having in mind the gauge invariant nature
and non-commutativity 
of $P_i$ or their analogs $\Pi_i$ from the extended
formulation, we conclude that they, like 
anyon coordinates $X_\mu$,
are the charge-monopole's analogs
of the Foldy-Wouthuysen  coordinates of the 
Dirac particle \cite{tor5,fw}.

\section{Concluding remarks}

The discussed classical $so(2,1)$ symmetry can be quantized
in an abstract way proceeding from the
classical relations (\ref{so21}) and (\ref{casim}).
In such a way the infinite-dimensional
unitary half-bounded $sl(2,R)$ representations
of the discrete series $D^+_{\alpha}$
will be obtained, which are characterized by 
the quantum Casimir element ${\cal C}=-\alpha(\alpha-1)$,
$0<\alpha\in\RR$ and by the eigenvalues $j_0=\alpha+n$,
$n=0,1,\ldots$, of the operator ${\cal J}_0$ \cite{sl2r}. 
Since $\alpha>0$ is arbitrary, such a quantization
procedure of the $so(2,1)$ symmetry algebra 
does not introduce any restrictions for the 
charge-monopole coupling constant,
and does not fix correctly the 
spectrum of the operator ${\bf J}^2$. 
The latter information, as we saw,
is encoded in the corresponding $so(3)$ algebra and 
classical condition ${\bf J}^2>\nu^2$.
The quantization of the parameter $\nu$ 
and the quantum spectrum of ${\bf J}^2$ 
could be obtained in principle by applying the geometric
quantization to the classical E(3) system
from section 2.4.
On the other hand, the same information
could be extracted from the quantization
of classical $so(3,1)$ symmetry of the charge-monopole
system described in section 2.3
with taking into account the classical relation 
${\bf J}^2>\nu^2$. 
We are going to consider the geometric quantization of $so(3,1)$ 
charge-monopole symmetry elsewhere.
The observation of the similarity between the charge-monopole 
and the 3D free particle systems realized in section 2.3 
in the context of the vector integrals of motion
has been applied recently in ref. \cite{fmsusy} 
for explaining the nature of the nonstandard 
fermion-monopole supersymmetry \cite{nonsusy}.

In sections 4 and 5 we have discussed
the two different interpretations of the 
charge-monopole system as a particle with
spin and as a spinning top system.
It is interesting to find  
the corresponding map between these pictures
at the Hamiltonian level.
In the first picture, the ``rotational" part
of the phase space is given by the 
spin vector ${\bf s}$, ${\bf s}^2=\nu^2$, and by
the orbital angular momentum ${\bf L}$
and associated unitary vector ${\bf n}$.
In the spinning top picture we have
the angular momentum vector ${\bf J}$ and the set of
orthonormal vectors ${\bf e}_a$.
The vector ${\bf J}$ of the second formulation
is identified with 
the total angular momentum vector
${\bf L}+{\bf s}$ from the first formulation,
and the vector ${\bf e}_3$ giving the
symmetry axis of the top is naturally
identified with ${\bf n}$.
Therefore, to find the mapping between the two formulations, 
it is sufficient to construct from the variables 
${\bf s}$, ${\bf L}$ and ${\bf n}$
the orthonormal vectors
${\bf e}_{1,2}$, ${\bf e}_1\times{\bf e}_2={\bf n}$,
satisfying the necessary Poisson bracket relations.
Unfortunately, we did not succeed in
realization of such a construction.

It seems interesting to
investigate analogously  other systems of particles
(strings) in the background of external gauge
or gravitational fields from the point of view
of their possible alternative description as the 
particles (strings) with internal reduced degrees 
of freedom. 

\vskip0.3cm
{\bf Acknowledgements}
\vskip3mm

I am grateful to J. Zanelli 
and D. Sorokin for interest to this work and helpful discussions.
I thank J. Alfaro and J. Gamboa for bringing 
refs. \cite{bbcel,godoli} to my attention and 
M. Ba\~nados for discussion of some related issues. 
The work has been supported by the
grant 1980619 from FONDECYT (Chile)
and by DICYT (USACH).

\end{document}